\documentclass[useAMS,usenatbib]{mn2e}

\usepackage{graphicx}
\newcommand{\be}{\begin{equation}}
\newcommand{\ee}{\end{equation}}
\newcommand{\ba}{\begin{align}}
\newcommand{\ea}{\end{align}}
\newcommand{\bea}{\begin{eqnarray}}
\newcommand{\eea}{\end{eqnarray}}
\newcommand{\rd}{\rm{d}}

\makeatletter
\def\elsartstyle{%
    \def\normalsize{\@setfontsize\normalsize\@xiipt{14.5}}
    \def\small{\@setfontsize\small\@xipt{13.6}}
    \let\footnotesize=\small
    \def\large{\@setfontsize\large\@xivpt{18}}
    \def\Large{\@setfontsize\Large\@xviipt{22}}
    \skip\@mpfootins = 18\p@ \@plus 2\p@
    \normalsize
}
\@ifundefined{square}{}{}
\makeatother

\begin{document}

\title{Deconstructing Baryon Acoustic Oscillations: \\A Comparison of Methods}
\author[Rassat et al.]{Ana\"is Rassat$^1$, Adam Amara$^2$, Luca Amendola$^3$, Francisco J. Castander$^4$,\and Thomas Kitching$^5$, Martin Kunz$^{6,7}$, Alexandre R\'efr\'egier$^1$, Yun Wang$^8$, Jochen Weller$^9$ \\
$^{1}$IRFU-SAP, Service d'Astrophysique, CEA-Saclay, F-91191 Gif sur Yvette Cedex, France.\\
$^2$Institute for Astronomy, ETH Hoenggerberg Campus, Physics Department, CH-8093 Z\"urich, Switzerland.\\
$^3$INAF - Osservatorio Astronomico di Roma, Via Frascati 33, 00040 Monte Porzio Catone, Roma, Italy.\\
$^4$Institut de Ci\`encies de l'Espai (IEEC/CSIC), Campus UAB, 08193 Bellaterra, Barcelona, Spain\\
$^5$University of Oxford, Department of Physics, Denys Wilkinson Building, Keble Road, Oxford, OX1 3RH, United Kingdom.\\
$^6$Institute for Theoretical Physics, University of Geneva, 24, Quai Ernest-Ansermet, CH-1211 Gen\`eve 4, Switzerland.\\
$^7$Department of Physics and Astronomy, University of Sussex, Brighton, East Sussex BN1 9QH, United Kingdom.\\
$^8$Dept. of Physics \& Astronomy, The University of Oklahoma, 440 W. Brooks St., Norman, OK 73019, USA.\\
$^9$Physics \& Astronomy, University College London, Gower Street, London WC1E 6BT, United Kingdom.}

\date{Accepted xxx. Received xxx; in original form xxx}

\maketitle

\begin{abstract}
The Baryon Acoustic Oscillations (BAOs) or \emph{baryon wiggles} which are present in the galaxy power spectrum at scales $100-150h^{-1}\rm{Mpc}$ are powerful features with which to constrain cosmology.  The potential of these probes is such that these are now included as primary science goals in the planning of several future galaxy surveys.  However, there is not a uniquely defined BAO Method in the literature but a range of implementations.  We study the assumptions and cosmological performances of three different BAO methods: the full Fourier space power spectrum [$P(k)$], the `wiggles only' in Fourier space and the spherical harmonics power spectrum [$C(\ell$)].  We contrast the power of each method to constrain cosmology for two fiducial surveys taken from the Dark Energy Task Force (DETF) report and equivalent to future ground and space based spectroscopic surveys.  We find that, depending on the assumptions used, the dark energy Figure of Merit (FoM) can change by up to a factor of 35 for a given fiducial model and survey.  We compare our results with the DETF implementation and, discuss the robustness of each probe, by quantifying the dependence of the FoM with the wavenumber range.   The more information used by a method, the higher its statistical performance, but the higher its sensitivity to systematics and implementations details.
\end{abstract}

\section{Introduction}\label{intro}
In the early Universe, just before recombination, fluctuations in the coupled baryon-photon fluid were subject to two competing effects: attractive gravity and repulsive pressure.  These two effects are expected to produce a series of acoustic peaks - dubbed Baryon Acoustic Oscillations (BAOs) - in both the Cosmic Microwave Background (CMB) \citep{Sunyaev:1970Z,Peebles:1970Y} and the matter power spectra \citep{Eisenstein:1999hu}.

These features have been observed in the CMB temperature-temperature power spectrum using data from several years of the Wilkinson Microwave Anisotropy Probe (WMAP) data  \citep[for the latest measurements see][]{Nolta:2008}.  As galaxy surveys cover increasingly larger volumes, they too can probe the scales on which the BAOs are predicted.  Recently these have been observed in the Sloan Digital Sky Survey (SDSS) \citep{Eisenstein:2005,Hutsi:2006bao,Percival:2007b} and the 2 degree Field Galaxy Survey (2dFGS) \citep{Percival:2007}.

The scale of the oscillations provides a standard ruler and has been used to constrain dark energy parameters \citep{Eisenstein:2005, Amendola:2005,Wang:2006M, wang:2006,Percival:2007b, Percival:2007,Ichikawa:2007T} as well as neutrinos masses \citep{Goobar:2006} and even alternative models of gravity \citep{Alam:2006,Pires:2006,Guo:2006,yamamoto:2006,wang:2007}, though current data does not cover enough cosmological volume to be constraining without the help of external data sets.  

In the future, BAOs will be a fundamental tool for precision cosmology \citep{Amendola:2005,WGFC,DETF} and there are many planned surveys which use BAOs as one of their primary science drivers.   These include ground based surveys such as the Dark Energy Survey \citep[DES]{DES:2005}, the Large Synoptic Survey Telescope \citep[LSST]{Zhan:2007K}, the Wide-Field Multi-Object Spectrograph \citep[WFMOS]{Glazebrook:2005}, WiggleZ \citep{Glazebrook:2007}, the Baryon Oscillations Spectroscopic Survey \citep[BOSS or SDSS III]{Schlegel:2007}, the Physics of the Accelerating Universe Survey \cite[PAU]{Benitez:2008}, the Hobby-Eberly Telescope Dark Energy Experiment \cite[HETDEX]{Hill:2004}, as well as space-based surveys such as the Advanced Dark Energy Physics Telescope (ADEPT),  the SPectroscopic All-sky Cosmic Explorer \citep[SPACE]{Cimatti:2008}, the Dark UNiverse Explorer \citep[DUNE]{Refregier:2008} and the EUCLID project\footnote{http://sci.esa.int/science-e/www/area/index.cfm?fareaid=102} currently under study by the European Space Agency.  Such surveys could probe the baryon oscillations in the galaxy distribution as well as the distribution of clusters \citep{Hutsi:2006}, quasars \citep{Schlegel:2007} or even supernovae \citep{BAOSNE:2008}.

One of the attractive properties of using the BAOs as a cosmological probe is that they are considered to represent a robust probe for extracting cosmological information. However there is not a unique BAO Method but a range of methods which differ in a number of ways. 

The first variable in the method is the statistics used in the analysis: the measurement can be done in real space \citep[$\xi(r)$]{Eisenstein:2005}, configuration space \citep[$w(\theta)$]{Loverde:2007bao}, Fourier space \citep[P(k)]{Seo:2003Eis,Amendola:2005} or in spherical harmonic space \citep[$C(\ell)$]{Dolney:2006}. Even when using the same statistic, there are approaches which use different levels of information and are therefore subject to different systematics.  For studies in Fourier space, some measure the full power spectrum information, including the BAOs \citep{Seo:2003Eis}, whereas others subtract the smooth part of the spectrum and focus only on the oscillations \citep{Blake:2006bao,Seo:2007Eis} - the latter uses less information but may be more robust with respect to systematic uncertainties.

In this paper we use the Fisher matrix approximation of the likelihood to contrast the information available in three of the methods described above: the Fourier space power spectrum [$P(k)$], the Fourier space `wiggles only' and the spherical harmonic correlation function [$C(\ell)$].

In section \ref{clustering} we overview the different features which are present in the galaxy power spectrum, and discuss the potential information carried by each feature. In section \ref{recipes} we describe the three BAO Methods used in this paper.  In section \ref{implementation} we describe the details of the Fisher forecast method used, as well as assumptions about our fiducial cosmological model and the implementations of each method.  In this section we also describe the ground and space based surveys corresponding to Stages III and IV for the BAO surveys described by the DETF report \citep{DETF}.  In section \ref{forecasts} we compare our results with the DETF implementation.  We then present the constraints derived for each method for both fiducial BAO surveys and we also combine these with CMB constraints derived from the future Planck mission.  In addition, we investigate the impact of the wavenumber range on the Figure of Merit for each BAO method.  In section \ref{discuss} we discuss the constraints obtained by each method, and the hierarchy that exists between them and give our overall conclusions.  Details of the implementation of the Fisher matrix calculations for each BAO method  as well as the for the Planck mission are given in Appendices \ref{app:forecast} and \ref{app:planck}. 

\section{The Building Blocks of the Galaxy Power Spectrum}\label{clustering}
\begin{figure*}
\includegraphics[height=12cm,angle=90]{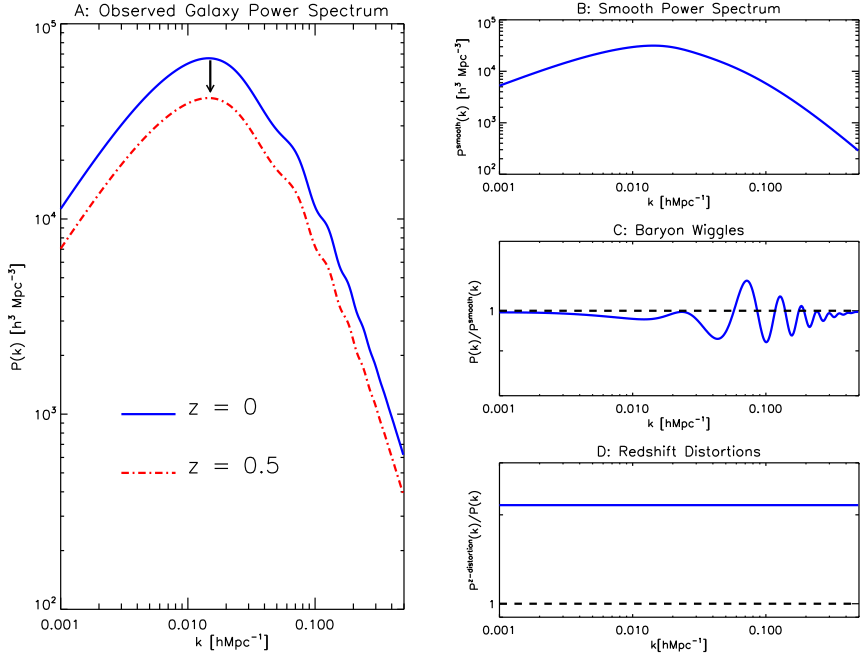}
\caption{Illustration of the main building blocks of the linear galaxy power spectrum.  {\bf Panel A}: The observed linear galaxy power spectrum.  This includes the linear galaxy bias, the smooth power spectrum (broad band power), baryonic wiggles and linear redshift distortions.  In linear theory the redshift evolution of the galaxy power spectrum depends solely on the linear growth factor (see equation \ref{clustering:eq:real}); we illustrate this by comparing the linear galaxy power spectrum at redshift $z=0$ (solid blue line) and redshift $z=0.5$ (dot-dashed red line).  {\bf Panels B, C and D} illustrate the different building blocks of the galaxy power spectrum.  {\bf Panel B: }The smooth part of the galaxy power spectrum contains information in the shape and normalization of the power spectrum; it can be calculated using the analytic formula given by Eisenstein \& Hu (1999). {\bf Panel C}:  The ratio (blue solid line) of the full power spectrum and the smooth part of the power spectrum reveals the residual baryonic wiggles.  The dashed line corresponds to no baryonic wiggles. {\bf Panel D:}  The ratio (blue solid line) of the radial galaxy power spectrum ($\mu = 1$ in equation \ref{clustering:eq:biasdef}) to the tangential spectrum ($\mu = 0$) illustrates the scale-independent effect of linear redshift distortions.  Linear redshift distortions add power in the radial direction.  The dashed line corresponds to no redshift distortions.}
\label{clustering:fig:pk}
\end{figure*}
The Fourier space matter power spectrum describes the fluctuation of the matter distribution and is defined by:
\be \left<\delta(\vec{k}) \delta^*(\vec{k'})\right>=(2\pi)^3\delta_D^3(\vec{k}-\vec{k'})P(k),\ee
where $\delta(\vec{k})$ represents the Fourier transform of the matter overdensities $\delta(\vec{r})=\frac{\rho(\vec{r})-\bar{\rho}}{\bar{\rho}},$
and the mean density of the Universe is $\bar{\rho}$. The term $\delta_D^3(\vec{k}-\vec{k}')$ represents the Dirac delta function.

The galaxy power spectrum is a rich statistic, where several features on different scales contain specific cosmological information.  On linear scales, we identify three main features in the galaxy power spectrum, which are:

\begin{itemize}
\item The broad-band power: Information is contained in the shape, normalization and time evolution of the power spectrum.
\item The baryon acoustic oscillations: Information is contained in the tangential and radial wavelengths, as well as the wiggle amplitude.
\item The linear redshift space distortions
\end{itemize}

The three building blocks of the observed galaxy power spectrum are plotted separately in Figure \ref{clustering:fig:pk} (panels B, C and D), while the observed power spectrum is plotted in panel A.  Each building block probes the various sectors of the cosmological model (dark energy, dark matter, initial conditions) in a different way.  In this section, we describe the main building blocks of the power spectrum and focus on their cosmological parameter dependence for the following 7 parameters: $[\Omega_m, \Omega_b, \Omega_{\rm DE}, w_0, w_a, h, n_s]$.

\subsection{Broad Band Power}\label{clustering:broad}

The initial dark matter power spectrum is assumed to be of the form: $P_{\rm {init}}(k) \propto k^{n_s}$. The spectral index $n_s$ controls the tilt of the initial power spectrum, and the Harrison-Zel'dovich value of $n_s=1$ corresponds to a scale invariant power spectrum of perturbations.  The linear matter power spectrum at a redshift $z$ can be obtained by assuming the initial power spectrum has evolved according to: 
\be P_{\rm{m}}(k, z) =D^2(z)P_{\rm{init}}(k)T^2(k).\label{clustering:eq:real}\ee
The quantity $T(k)$ is the transfer function and is a solution to Boltzman's equations which includes the baryon wiggles.  The linear growth factor $D(z)$, normalized to $1$ today, quantifies the evolution of linear growth of structure with redshift.  Both the matter $\Omega_m$ and the dark energy $\Omega_{\rm DE}$ densities affect the normalization of the power spectrum.  

The linear power spectrum today is characterised by a turnover (which occurs around $k_{eq} \sim 0.015h^{-1}\rm{Mpc}$ in Figure \ref{clustering:fig:pk}); the turnover corresponds to the size of modes which entered the horizon at matter-radiation equality \citep{Dodelson:2003}. The scale $k_{\rm eq}$ at which this occurs is proportional to $\Omega_m h^2$ (with no dependence on the radiation density, since this is well constrained by the measurements of the CMB temperature anisotropies).  

In a flat universe, the dark energy density $\Omega_{\rm DE}$ will indirectly affect the position of the turnover through $\Omega_{\rm DE} = 1 - \Omega_m$.  In this work we always consider a curved universe where $\Omega_{\rm DE}$ and $\Omega_m$ are independent of each other, so that the dark energy density does not affect the position of the turnover but it does affect the overall amplitude.  In the case where dark energy clusters (i.e. its equation of state parameters are given by $w_0 \neq -1$ and $w_a \neq 0$ when using the parameterisation defined in equation \ref{implementation:eq:wo_wa}), the power spectrum will be enhanced on large scales \citep{Ma:1999cbw}.

 Increasing the baryon density, $\Omega_b$, will have the effect of decreasing power on small scales as well as also introducing baryon wiggles in the power spectrum (see section \ref{clustering:bao}).  In this paper, the matter density is the sum of the Cold Dark Matter (CDM) and baryon (b) densities respectively, i.e. $\Omega_m = \Omega_{\rm CDM} + \Omega_b$, where we choose $\Omega_m$ and $\Omega_b$ to be independent.  This means that when increasing $\Omega_b$, $\Omega_{\rm CDM}$ must decrease in order for $\Omega_m$ to be kept fixed.

In a galaxy survey, the observable is the galaxy overdensity $\delta_{\rm g}$ which is assumed to trace the underlying matter distribution $\delta_m$ following:
\be \delta_{\rm g} = b(z, k) \delta_m, \ee
where $b(z,k)$ is the galaxy bias, which can be both redshift and scale dependent. The real space galaxy power spectrum is then given by
\be P_{\rm{g}}(k, z) =b^2(z, k)D^2(z)P_{\rm{init}}(k)T^2(k).\ee

On linear scales, it is often assumed that the scale dependence of the galaxy bias can be dropped so that $b(k,z)=b(z)$; in this case, the linear bias only modulates the overall amplitude of the galaxy power spectrum $P_{\rm g}(k)$.  The bias at redshift zero [$b(z=0)$] is degenerate with the normalization $\sigma_8$, so that in general the bias $b(z)$ is degenerate with both the normalization and the growth $\sigma_8\cdot D(z)$.

\subsection{Baryon Acoustic Oscillations}\label{clustering:bao}

The baryon acoustic oscillations in the CDM power spectrum are relatively weak features (see Figure \ref{clustering:fig:pk}); nevertheless they can be used independently to constrain cosmology \citep{WGFC,DETF}.  They can be isolated from the galaxy power spectrum either by taking the ratio with the corresponding
baryon-free power spectrum, or a smooth fitting curve \citep{Blake:2003G,Seo:2003Eis}.
In the latter case, the oscillations can be
approximated by a decaying sinusoidal function. The amplitude
of the oscillations increases with the baryon density $\Omega_b$,
and the location $k_A$ of the peaks is related to the sound horizon
at decoupling, $s$, by $k_A = 2\pi/s$ \citep{Blake:2003G}.

The theoretical value of $k_A$ provides a known - or \emph{standard} - ruler,
fixed by the sound horizon at decoupling:
\be s = \int_0^{t_{\rm dec}} \frac{c_s}{a} {\rm d}t
= \frac{c}{H_0} \int_{z_{\rm dec}}^\infty \frac{c_s}{E(z)} \rd z, \ee
where $c_s$ is the sound speed and: \be E(z)=H(z)/H_0=\sqrt{\Omega_{\rm rad}(1+z)^4+\Omega_m(1+z)^3+\Omega_{\rm DE}f(z)},\label{clustering:bao:eq:ez}\ee where the expression \be f(z)=\exp\left[3\int_0^z \frac{ 1+w(z')}{1+z'}dz'\right],\ee describes the effect of dark energy on the Universe's expansion.  At these high redshifts the curvature can be neglected. The radiation density is strongly constrained by the measurement of the CMB temperature, and the sound horizon $s$
depends strongly on $\Omega_m$ through equation \ref{clustering:bao:eq:ez}.  In addition, the sound speed and the redshift of decoupling $z_{\rm dec}$ depend on $\Omega_bh^2$, which is thereby moderately constrained, but apart from this combination, the constraints on $\Omega_b$ and $H_0$ \emph{separately} are weak due to degeneracies (though including the amplitude of the baryon wiggles would further constrain $\Omega_b$).

Dependence on dark energy parameters
can vary with the model, but in most cases the dark energy density
at very early times is small and can be neglected (\cite{Bean:2001} showed it has to be
less than a few percent during the Big Bang Nucleosynthesis epoch).  The dependence on the dark energy parameters does therefore not enter
through the size of the ruler, it affects the \emph{observed} oscillation scale:
A ruler at a redshift $z$ with a given (comoving) size $r_\perp(z)$
perpendicular to the line of sight, and a size $r_{\parallel}(z)$ along
the line of sight is related to the observed angular size $\Delta\theta$
and redshift extent $\Delta z$ through \citep{Seo:2003Eis}: \begin{eqnarray}
r_\perp(z) &=& (1+z) D_A(z) \Delta\theta, \\
r_{\parallel}(z)  &=& \frac{c\Delta z}{H(z)},\end{eqnarray}
where $D_A(z)$ is the angular diameter distance.  Since the size of the ruler scales with $s$, the observation of the
baryon acoustic scale in the tangential and radial direction affords us a
measurement of $D_A/s$ and $H \cdot s$ respectively. In the limit where
the size of the sound horizon is known from CMB data to
much higher precision than the measurement of the oscillation
scales from the galaxy survey, we can consider the baryon wiggles
to provide estimates of the angular diameter distance and the
expansion rate directly.

Thus, if we divide out the broad shape of the power spectrum and
neglect the amplitude of the wiggles, we expect to be able to
measure the matter and dark energy densities $\Omega_m$ and
$\Omega_{\rm DE}$; with sufficient amount of redshift information, it will also be possible to constrain the evolution of the dark energy equation of state $w(z)$.

\subsection{Redshift Space Distortion}\label{clustering:zspace}

An observer can only measure the galaxy power spectrum in redshift space, which is distorted compared to the power spectrum in real space. Redshift distortions are due to peculiar velocities of galaxies; these cause radial distortions in the observed galaxy density field.  This distortion occurs because the observed redshift $z_{obs}$ is a sum of two quantities: \be z_{obs} = z_h + z_{\vec{v} . \vec{r}},\ee where $z_h$ is the redshift due to the cosmological Hubble expansion of the Universe, and $z_{\vec{v} . \vec{r}}$ is the redshift due to the radial component of the galaxy's peculiar velocity.  

In linear theory this will affect structures along the line of sight, which will appear enhanced compared to transverse structures; i.e. for a structure which is isotropic in real space, an observer will measure more power in the radial direction than in the transverse direction. Considering linear theory alone, the observed redshift space power spectrum for galaxies is related to the real space matter power spectrum by \citep{Kaiser:1987,Seo:2003Eis}: \be P^{g,z}(k, z, \mu) \propto \left(1+\beta \mu ^2\right)^2b(z,k)^2P^{\rm {m,real}}(k, z), \label{clustering:eq:biasdef}\ee where $\mu$ is the cosine of the angle between the wavevector $\vec{k}$ and the line of sight.  The amplitude of the redshift distortion is modulated by the distortion parameter \be \beta=\frac{1}{b(z)}\frac{{\rm dln}D(a)}{{\rm dln}a},\ee so that the redshift distortions are a probe of the growth rate of structure as well as the linear galaxy bias.

\section{Recipes for Baryon Acoustic Oscillations}\label{recipes}

\begin{table*}
\caption{Schematic summary of which building blocks of the galaxy power spectrum are probed by each family of BAO method considered in this paper.  In this paper, certain components such as the normalization of the power spectrum or the redshift distortions have been marginalized over in order to follow prescriptions which exist in the literature; when this is done the corresponding building block is represented in brackets in the Table below.}

\begin{center}
\begin{tabular}{|l|c|c|c|c|c|c|}
\hline
&Broad Band&Overall Amplitude&Tangential&Radial& Wiggle Amplitude&Redshift Space\\
&Power&&BAO scale&BAO scale&&Distortions\\
\hline
\hline 
Full $P(k)$&$\surd$&$\left(\surd\right)$&$\surd$&$\surd$&$\surd$&($\surd$)\\
`Wiggles only'&--&--&$\surd$&$\surd$&--&--\\
$C(\ell)$&$\surd$&$\left(\surd\right)$&$\surd$&--&$\surd$&$\surd$\\
\hline
\end{tabular}
\end{center}
\label{recipes:tab:summary}
\end{table*}%

The goal of this paper is to contrast the constraining potential of three different families of BAO methods, namely:
\begin{enumerate}
\item Full Fourier space galaxy correlation function $P(k)$,
\item Fourier space BAO `wiggles only',
\item Spherical harmonic space galaxy correlation function $C(\ell)$.
\end{enumerate}

The general underlying basis of each method is described in this section and a schematic summary is given in Table \ref{recipes:tab:summary}.  Details of the tomographic method, the fiducial cosmological model and further implementation specifications are given in section \ref{implementation}.  Full mathematical details on the Fisher matrix calculations for each family of methods are given in Appendix \ref{app:forecast}.

\subsection{Full P(k)}
The full Fourier space galaxy correlation method uses information across different scales and capitalizes on each building block of the galaxy power spectrum. Such methods have already been applied to forecasts for future surveys \citep[see Appendix \ref{app:forecast:pk}]{Seo:2003Eis,Amendola:2005}.  

In this paper we follow the implementation of \cite{Seo:2003Eis}, which we refer the reader to for specific methodology, though details are also given in section \ref{implementation} and Appendix \ref{app:forecast:pk}.

Intuitively, because this method uses information from the full galaxy correlation function, it should have the potential to constrain cosmological parameters with high precision.  However since it uses information over a wide range of scales this method could also be prone to a high level of systematics.  

In particular, the unknown linear bias $b(z)$ will affect both the overall amplitude of the power spectrum as well as the amplitude of the redshift distortion (see equation \ref{clustering:eq:biasdef}).  If the bias is also scale dependent, i.e. if $b(z)=b(k,z)$, then this can potentially distort the shape of the power spectrum over a range of scales leading to a fairly high sensitivity to systematics.  

The full galaxy correlation function will also be more sensitive to non-linear redshift distortions at small scales.  Additionally, intrinsic non-linearities in the density contrast will distort the shape of the spectrum at large wavenumbers - while leaving the wiggle location relatively unchanged (see for e.g. \cite{scoccimarro:2004} and \cite{Matarrese:2008}). 

\subsection{BAO wiggles only}
To avoid the effect of the potential systematics of the full $P(k)$ method, one can focus the analysis on specific scales, and remove the overall shape and amplitude of the galaxy power spectrum, centering the analysis on the baryon wiggles only. 

The way to do this is to consider the ratio of the observed power spectrum to a baryon-free power spectrum, i.e. $P(k, \Omega_b \ne 0)/P(k, \Omega_b=0)$ (as in Figure \ref{clustering:fig:pk}) or to a smooth parametric curve \citep{Blake:2006bao,Seo:2007Eis}.  This BAO `wiggle-only' method has been implemented in several papers. Some popular methods, are described in \cite{Parkinson:2007}, \cite{Blake:2006bao} and \cite{Seo:2007Eis}.  

By focusing on the baryon wiggles only, the measured quantity is now independent of the redshift dependent linear bias $b(z)$ (though the errors on the peak measurement will depend weakly on the bias) and, because the wiggles occur on a limited wavenumber range, the measured quantity is also weakly dependent on the scale dependent bias $b(k)$.  This method should intuitively be more robust than the full $P(k)$ approach, as it exploits less features over a limited $k$ range, it is also bound to provide weaker constraints on the cosmological parameters.

In this paper, the main calculations for the `wiggles only' method are performed using the formalism of \cite{Seo:2007Eis} and \cite{Parkinson:2007} (see section \ref{implementation} and Appendix \ref{app:forecast:bao}), except in section \ref{forecasts:detf} where the errors on the wiggles estimation are calculated as in \cite{Blake:2006bao}.  

\subsection{Spherical Harmonics $C({\ell})$}
Both methods described above perform the analysis in Fourier space.  Doing this raises several issues.  

The first is that equation (\ref{clustering:eq:biasdef}) which relates the real and redshift space Fourier power spectra is only valid in the far field approximation, i.e. for galaxies which are separated by small angles on the sky, and for which the line of sight vectors can be considered parallel.  While this may be a valid approximation for current surveys, future surveys plan to cover all the extragalactic sky ($\sim 2\pi$) and this may no longer be valid.  A natural decomposition for data on a sphere is in spherical harmonic space in which transverse and radial modes are independent.  In this decomposition, there exists an exact solution to express the galaxy power spectrum in redshift space \citep[see][and equations \ref{app:forecast:cl:eq:cl} and \ref{app:forecast:cl:eq:clredshift} in Appendix \ref{app:forecast}]{Fisher:1994,Heavens:1995}.  For large sky coverage it is therefore more judicious to study the galaxy power spectrum in spherical harmonic space.

The second and more fundamental issue has to do with the actual measurement of the Fourier power spectrum.  The bare observables provided by a redshift survey are the angular position and redshift of a galaxy, say $(\theta, \phi, z)$ (although technically the redshift is itself a first order quantity derived from the galaxy's observed electromagnetic spectrum).  To measure the Fourier space power spectrum it is necessary to relate these observables to the wavenumber $\vec{k}$ and this transformation is dependent on the assumed cosmological model.  On the other hand, in spherical harmonic space, the transform from ($\theta, \phi, z$) to ($\ell, z$) is independent of cosmology and this is a further central motivation for analysing galaxy surveys in spherical harmonic space.  

This is the final family of methods we investigate, following the tomographic method described in \cite{Dolney:2006} (see section \ref{implementation} and Appendix \ref{app:forecast:cl} for details).  

\section{Implementation Details}\label{implementation}
In this section we describe the implementation details of our calculations.  In section \ref{implementation:cosmology}, we justify our choice of fiducial cosmological parameters and their central values.  In section \ref{implementation:survey} we describe the two fiducial surveys we use for our calculations, which are chosen as Stage III and Stage IV BAO surveys as described in the DETF report \citep{DETF}; these correspond to future ground and space based surveys respectively.  In section \ref{implementation:details}, we overview the specifications of our calculations.  

\subsection{Fiducial cosmological model and central values}\label{implementation:cosmology}

To quantify the dark energy, we adopt a convenient parameterisation \citep{Chevallier:2001, Linder:2003} of its equation of state $w(a)$ by,
\be w(a)=w_0 +(1-a) w_a,\label{implementation:eq:wo_wa}\ee where $a=1/(1+z)$.  This can also be expressed in terms of the pivot scale factor $a_p$: \be w(a) = w_p + (a_p-a) w_a.\ee  The DETF ``Figure of Merit'' (FoM) is defined by:
\be {\rm FoM} = \frac{1}{\sigma(w_p)\sigma(w_a)},\label{implementation:eq:fom}\ee which is used to compare the suitability of different probes and surveys to constrain the dark energy equation of state.  In order to avoid problems of phantom crossing around $w_0=-1$, we chose a central value of $w_0=-0.95$.

We forecast errors for an 7-parameter model within the framework of general relativity, i.e. we have not added any extra parameters to account for potential modified gravity scenarios \citep{Amendola:2008Kunz,Huterer:2007lin}. 
 The 7-parameter model and central fiducial values are given in Table \ref{implementation:tab:central}.  The power spectrum normalization is chosen as $\sigma_8=0.80$ and is marginalised over for calculations involving the full Fourier power spectrum and the spherical harmonic galaxy correlation function.

\subsection{Fiducial Survey}\label{implementation:survey}

The DETF have published dark energy forecasts for different ``Stages'' of surveys, numbered from I to IV.  Stage I corresponds to dark energy projects that have already been completed; Stage II to ongoing projects, Stage III to ``near-term, medium-cost, currently proposed projects'' and Stage IV to future ground or space based (LSST/SKA or JDEM like) missions with large sky coverage.  The BAO forecasts published by the DETF were calculated using a specific implementation of the `wiggles only' method.  In this paper we are interested in comparing these forecasts with those from two other methods, namely the full Fourier power spectrum and the spherical harmonic power spectrum. 

\begin{table}
\caption{Central fiducial values and cosmological parameter set for the Fisher matrix calculations performed in section \ref{forecasts}.  The normalization of the power spectrum is taken as $\sigma_8$=0.80 .}
\begin{center}
\begin{tabular}{|l|c|}
\hline
Parameter&Central fiducial value\\
\hline 
\hline
$\Omega_m$&0.25\\
$\Omega_b$&0.0445\\
$\Omega_{\rm DE}$&0.75\\
$w_0$&-0.95\\
$w_a$&0\\
$h$&0.70\\
$n_s$&1\\
\hline
\end{tabular}
\end{center}
\label{implementation:tab:central}
\end{table}%

\begin{table*}
\caption{Future fiducial spectroscopic ground and space based surveys taken from DETF.  The sky coverage, redshift range and galaxy density ($nP = 3$) are described in the DETF report.  The number of redshift bins, the bias prescription and wavenumber range are chosen specifically for this paper.}
\begin{center}
\begin{tabular}{|l|l|l|}
\hline
&Ground Based Survey&Space Based Survey\\
\hline 
\hline
DETF denomination&BAO-IIIS-o (WFMOS-like, ``wide'' only)&BAO-IVS-o (Optical/NIR JDEM Spatial Mission)\\
Sky coverage&$2,000 {\rm deg}^2$ ($f_{\rm sky} = 0.05$)&$10,000 {\rm deg}^2$($f_{\rm sky}=0.25$)\\
Redshift range&$0.5<z<1.3$ (4 bins)&$0.5<z<2$ (8 bins)\\
\hline
Bias prescription&\multicolumn{2}{c}{$b_{\rm g}(z) = \sqrt{1+z}$~~~~~~~~~~~~~~~~~~~~~}\\
$k$-range& \multicolumn{2}{c}{$k_{{\rm min}} = 10^{-3}$, $k_{{\rm max}}<0.25$ or $\sigma(R)<0.20$~~~~}\\
\hline
\end{tabular}
\end{center}
\label{implementation:tab:stagesIII_IV}
\end{table*}%
\begin{figure}
\includegraphics[width=8cm]{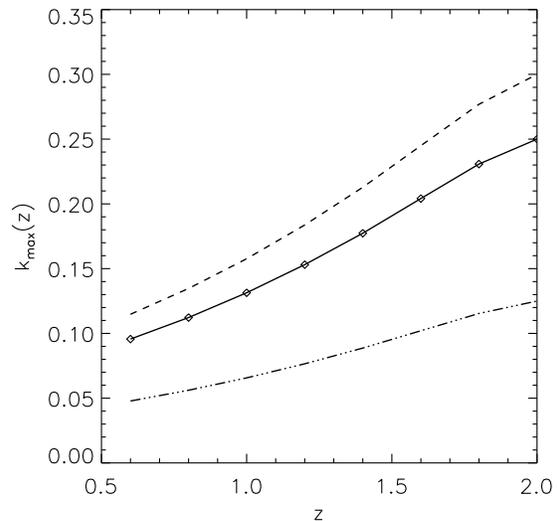}

\caption{The value of $k_{\rm max}$ as a function of redshift z.  The solid line (diamonds) corresponds $\sigma(R)<0.20$ and $k_{\rm max}<0.25 h{\rm Mpc}^{-1}$.  The dashed and dot-dashed lines correspond respectively to deviations around the central line, specifically $k_{\rm max}(z)\cdot 0.5$ (dot-dashed) and $k_{\rm max}(z)\cdot 1.2$ (dashed) respectively and are used in section \ref{forecasts:kmax}.}
\label{implementation:fig:kmax}
\end{figure}

We make these comparisons for future ground and space based spectroscopic (``S'' identifier in DETF denomination) surveys, which correspond to ``Stage III'' and ``Stages IV'' surveys.  In the DETF report, the spectroscopic Stage III survey is composed of both a wide and a deep survey (BAO-IIIS-o in DETF denomination, defined page 56 in \cite{DETF}), for which results are always presented combined.  Here we consider only the wide survey of the Stage III.  For the Stage IV mission, we chose it to resemble a spectroscopic JDEM-like mission, by using the DETF notation BAO-IVS-o (defined page 56 in \cite{DETF} - the optimistic survey does not include systematics).  

The main features of these two spectroscopic surveys are given in Table \ref{implementation:tab:stagesIII_IV}.  The galaxy number density was taken so that $n_{\rm eff}P_{\rm g}(k^*=0.2h{\rm Mpc}^{-1})=3$ \citep{Blake:2006bao, DETF}, which fixes the galaxy distribution and shot noise contribution.  We note that fixing the value of $n_{\rm eff}P(k^*)$ does not construct a realistic magnitude-limited galaxy distribution. The sky coverage and redshift ranges are chosen according to the description given in the DETF report and given in Table \ref{implementation:tab:stagesIII_IV}.  Specifications given in the second half of the Table are chosen specifically for this paper and are explained in more detail below.

\subsection{Specifications}\label{implementation:details}

For the `wiggles only' calculation (which is the calculation performed by the DETF) the value of the galaxy bias is irrelevant - so long as it is scale independent, which is what we assume here.  For the Fourier space and spherical harmonic space calculations, we chose the bias to follow a simple ad hoc functional form of $b(z) = \sqrt{1+z}$.  This choice can potentially affect statistical constraints for the full $P(k)$ and $C(\ell)$ methods, where the bias is explicitly present; whereas the `wiggles only' method depends very weakly on the value of the galaxy bias.

As the linear bias and normalization $\sigma_8$ are highly degenerate, we also marginalise over the bias value $b(z_i)$ in each redshift bin as well as over $\sigma_8$ for the $C(\ell)$ method (which is slightly different than the \cite{Dolney:2006} implementation, who uses a more complex bias prescription).  

For the full $P(k)$ method we follow the implementation of \cite{Seo:2003Eis}: information from the amplitude of the power spectrum as well as from redshift distortions are suppressed by marginalizing over the product $\sigma_8 b(z_i)D(z_i)$ and the redshift distortion parameter $\beta(z_i)$ in each redshift bin.  Finally, we marginalize over an unknown white shot noise, which is denoted in \cite{Seo:2003Eis} by $P_{\rm shot}$.

For all methods we perform calculations in the linear regime.  This is ensured by restricting the calculation to certain linear scales. The $k_{\rm max}$ at which non-linear effects become no longer negligible in
our fiducial model depends of course on the precision one wishes to
attain. Recent work \citep{Crocce:2008, Matarrese:2008} show that at $k=0.2 h {\rm Mpc}^{-1}$ the deviation from linearity at z=0 is already
sizeable, although the position of the peaks is much less sensitive to
non-linear corrections.  

The wavenumber corresponding to the linear cut-off will evolve with redshift. We quantify this evolution by considering only scales for which $\sigma(R)<0.20$ and $k_{\rm max}<0.25 h{\rm Mpc}^{-1}$, where $\sigma(R)$ is defined similarly to the normalization $\sigma_8\equiv \sigma(R=8 h^{-1} \rm {Mpc})$, but for a general $R$.  The evolution of the $k_{\rm max}$ we use for our computation with redshift bin is illustrated in Figure \ref{implementation:fig:kmax} as the solid line with diamonds.  We assume the largest scales probed are given by $k_{\rm min}=10^{-3}h{\rm Mpc}^{-1}$.

We find the choice of wavenumber range can have a very strong influence on results.  We investigate the dependence of the FoM on the values of $k_{\rm max}$ in section \ref{forecasts:kmax}.

\section{Forecasts}\label{forecasts}
The `wiggles only' BAO method we use is comparable to that used by the Dark Energy Task Force \cite[DETF]{DETF}, and so we first compare constraints from our calculation with those of DETFast\footnote{http://www.physics.ucdavis.edu/DETFast/ by Jason Dick and Lloyd Knox.} in section \ref{forecasts:detf}.  In section \ref{forecasts:compare}, we compare the constraining potential of the three different families of BAO methods.  The constraints are calculated for the galaxy surveys described in section \ref{implementation}, namely a future ground- and space-based spectroscopic surveys.

\subsection{Comparison with DETF Calculation}\label{forecasts:detf}
In this paper, constraints for the BAO `wiggles only' method are performed using distance error estimates from \cite{Seo:2007Eis}.  The DETF uses a similar method, though the distance error estimates are calculated using the fitting formulae given by \cite{Blake:2006bao}.  The slight difference between the fitting formulae provided in \cite{Blake:2006bao} is that the $k$-range is fixed, whereas in \cite{Seo:2007Eis} the $k$-range is an input to the error estimates.  In this section we switch to using the error estimates from \cite{Blake:2006bao} in order to compare constraints with those given by the DETF.  We obtain constraints from DETFast\footnotemark[1], a web-interface for calculations performed by the DETF.

In this calculation the central fiducial model and cosmological parameter set was changed slightly to adopt the same as that in the DETF report, i.e. $\Omega_{DE}=0.7222$, $\Omega_k=0$, $w_0=-1$, $w_a=0$, and $\Omega_m h^2=0.146$ and the sound horizon is kept fixed.  The fiducial survey was chosen to be BAO-IVS-o, as described in Table \ref{implementation:tab:stagesIII_IV}.  Comparison between constraints is given in Table \ref{forecasts:tab:detf}; the reported constraints are shown without any priors (i.e. without Planck priors).  Both calculations show agreement at the sub-percent level for all marginalised errors (as well as for fixed errors - not shown in this table).

\subsection{Comparison of different BAO Methods}\label{forecasts:compare}

\begin{table*}
\caption{Marginalised 1$\sigma$ constraints from the three different BAO Methods for DETF like ground-based survey (BAO-IIIS-o) and space-based survey (BAO-IVS-o).}
\begin{center}
\begin{tabular}{|c|l|l|l|l|l|l|}
\hline
&\multicolumn{2}{l}{Fourier space BAO `wiggles only'}&\multicolumn{2}{l}{Spherical Harmonic Space $C(\ell)$}&\multicolumn{2}{l}{Fourier space Full $P(k)$}\\
\hline 
\hline
Ground Based Survey (with Planck priors)&&\\
\hline
$\Omega_m$&2.04& (0.170)&0.303 &(0.162)&0.137 &(0.01)\\
$\Omega_b$&$>10$ &(0.0303)&0.122 &(0.0290)&0.0310 &(0.002)\\
$\Omega_{\rm DE}$&1.99& (0.156)&1.25&(0.163)&0.215 &(0.016)\\
$w_0$&6.28 &(2.42)&3.23&(1.69)&0.56 &(0.25)\\
$w_p$&1.15 &(0.184)&0.979&(0.650)&0.358 &(0.07)\\
$w_a$&$>10$ &(5.41)&$>10$&(4.45)&3.07& (0.65)\\
$h$& $>10$ &(0.238)&2.38&(0.227)&0.13 &(0.016)\\
$n_s$&-&(0.00465)&1.54&(0.00465)&0.165 &(0.0045)\\
\hline
Dark energy FoM&$<0.090$ &(1.2)&$<0.10$&(0.35)&0.91 &(21)\\
\hline
Space Based Survey (with Planck priors)\\
\hline 
$\Omega_m$&0.181 &(0.0385)&0.0367&(0.0254)&0.00985 &(0.003)\\
$\Omega_b$&$>10$ &(0.00687)&0.0193&(0.00463)&0.00221& (0.0006)\\
$\Omega_{\rm DE}$&0.178& (0.0375)&0.113&(0.0251)&0.0692& (0.005)\\
$w_0$&0.990 &(0.377)&0.355&(0.331)&0.0638 &(0.049)\\
$w_p$&0.197 &(0.0410)&0.147&(0.0955)&0.0313& (0.02)\\
$w_a$&3.94 &(0.878)&1.63&(0.610)&0.612 &(0.10)\\
$h$& $>10$ &(0.0539)&0.244&(0.0360)&0.0118 &(0.004)\\
$n_s$&- &(0.00463)&0.182&(0.00450)&0.0106 &(0.0034)\\
\hline
Dark energy FoM&1.5 &(28)&4.2&(17)&52& (502)\\
\hline
\end{tabular}
\end{center}
\label{forecasts:tab:constraints}
\end{table*}%

For the comparison between the three different BAO methods, we switch back to the fiducial cosmological model given in Table \ref{implementation:tab:central}.  We also use error estimates from \cite{Seo:2007Eis} for the `wiggles only' method, in order to have control over the $k$-range.  

The three methods are: the full power spectrum in Fourier space $[P(k)]$, the Fourier space wiggles only method [BAO `wiggles only'] and the full projected power spectrum in spherical harmonic space $[C(\ell)]$.  In Table \ref{forecasts:tab:constraints} we present the marginalised cosmological constraints for the 7-parameter model we consider from these three different BAO methods.  These constraints are marginalised over the linear bias and power spectrum normalization ($\sigma_8$) for the $C(\ell)$ method; for the full $P(k)$ method we marginalise over the amplitude of the power spectrum and the redshift distortions term as detailed in section \ref{implementation:details}.

The results in brackets have been combined with forecasted constraints from the Planck CMB mission, or what we refer to from now on as Planck priors.  The forecasted CMB priors include scalar temperature and E-mode polarization information.  Details on how we calculated the Planck priors are given in Appendix \ref{app:planck}.  

\begin{figure}
\includegraphics[width=8cm]{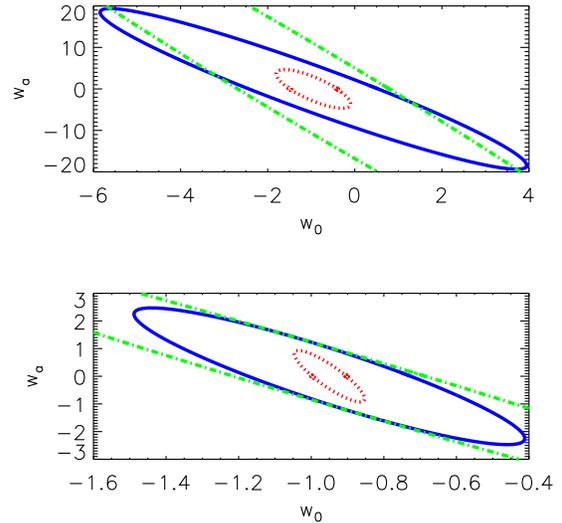}
\caption{Marginalised 1$\sigma$ constraints (without Planck priors) on dark energy equation of state parameters $w_0$ and $w_a$ for the three different BAO methods: Red (dotted): Full P(k), Blue (solid): C(l), Green (hashed: BAO `wiggles only'.  TOP: ground based survey (BAO-IIIS-o). BOTTOM: space-based survey (BAO-IVS-o). }\label{forecasts:fig:contours}

\end{figure}

Parameters for which the table entry shows `$-$', are not constrained by the corresponding method (i.e. the affected parameter is omitted from the Fisher matrix calculation), and we have not quantified constraints which were weaker than $10$.  The cosmological constraints for different methods are presented in Figure \ref{forecasts:fig:contours} for the dark energy parameters $w_0$ and $w_a$ for the ground (top panel) and space (lower panel), without Planck priors.

We first explore the results from Table \ref{forecasts:tab:constraints} without Planck priors as this gives us an understanding of what each BAO method constrains.  

We begin by exploring the ground-based survey without Planck priors.  The `wiggles only' method, which uses no information from the broad band power, the redshift space distortions nor the wiggle amplitude (see Table \ref{recipes:tab:summary}), gives constraints which are up to a factor of $7$ weaker than the $C(\ell)$ method.   Furthermore, the `wiggles only' method fails to constrain either $\Omega_b$, $w_a$, $h$ and $n_s$, whereas the $C(\ell)$ method only fails to constrain $w_a$.  The constraints from the $C(\ell)$ method are a factor 2 to 18 weaker than those from the $P(k)$ method, which can constrain $w_a$ (though the errors are still large).  In all cases, the Figure of Merits (FoMs) are small, making any comparison between them meaningless (for very low FoMs, upper limits only are shown in Table \ref{forecasts:tab:constraints}).

Adding the Planck priors to the ground-based case improves constraints (hereafter we omit improvements on previously unconstrained parameters) from the `wiggles only' method by a factor of $3-13$, from the $C(\ell)$ method by a factor of $1.5-10$ (omitting the improvement by a factor of $300$ on $n_s$), from the $P(k)$ method by a factor of $2-37$.  Including the Planck priors also reduces the differences between the `wiggles only' and the $C(\ell)$ methods (the marginalised constraints are of the same order of magnitude for all parameters), but the constraints from the $P(k)$ method are for some parameters up to a factor 16 tighter than for the $C(\ell)$ method.  The FoM for the $P(k)$ method is over a factor of $17$ larger than the FoM for both the `wiggles only' and $C(\ell)$ method.

We now consider constraints for the space-based survey, beginning with the constraints without priors.

For the space-based survey, constraints from the `wiggles only' method are a factor of $1.5-5$ weaker than those for the $C(\ell)$ method, and $\Omega_b$ and $h$ still remain unconstrained for the `wiggles only' method (which is expected, since this method constrains only the product $\Omega_mh^2$, see \ref{clustering:bao}).

The constraints from the $P(k)$ method remain considerably stronger, giving an improvement of $2-21$ on both the `wiggles only' and the $C(\ell)$ method, where the largest improvement (factor of improvement greater than 15) is on $h$, $n_s$, $\Omega_m$ and $w_0$.  The FoM for the $P(k)$ method is 52, i.e., 35 times larger than that for the `wiggles only' method and 12 times larger than that for the $C(\ell)$ method.

Adding the Planck priors to the space-based case improves constraints from the `wiggles only' method by a factor of $3-5$, on the $C(\ell)$ method by a factor of $1.5-7$ (omitting the factor of $40$ improvement on $n_s$), and on the $P(k)$ method by a factor of $1.3-14$.  As with the ground-based survey, the space-based survey with the Planck prior shows little variation between the `wiggles only' and the $C(\ell)$ methods, with the Figure of Merit for each method being the same order of magnitude.  The FoM for the $P(k)$ with Planck priors is 502, approximately 18 times larger than the FoM for the `wiggles only' method with Planck priors, and  a factor of 30 tighter than $C(\ell)$' method with Planck priors.

The constraints for $w_0$, $w_a$, $\Omega_{\rm DE}$ and $w_p$ for the space-based survey, using the `wiggles only' method with Planck priors are of the same order of magnitude of those presented by the DETF for BAO-IVS-o + Planck (page 77 of \cite{DETF}), though comparison should be taken with caution as both use different calculations for the Planck prior as well as different methods of calculating the errors for the BAO calculation.  The DETF report does not quote constraints on the other parameters, though they have also been marginalised over.

In order to investigate which building block of the power spectrum provides the most information to the $P(k)$ method, we evaluate new constraints, this time for a quasi baryon-free spectrum (to do this we set $\Omega_b=0.005$ as lower values causes numerical problems as other authors have also found).  This test suppresses the baryon wiggles, leaving the broad band power as the only building block which now provides information, since the amplitude and the redshift distortions are marginalised over.  

For the space-based survey, we find a new FoM of 11 (123) without (with) Planck priors, compared to the previous values of 52 (502) for the calculation including the baryon wiggles.  Though the values of the FoM for the $P(k)$ method have dropped by about a factor of 4 when omitting information from the baryon wiggles, these constraints are still an order of magnitude tighter than those for the `wiggles only' method, suggesting that a large part of the constraining power of the $P(k)$ method comes from the information contained in the broad band power.
\begin{table}
\caption{Comparison of 1$\sigma$ marginalised constraints from BAO `wiggles only' (without Planck priors with the DETFast results.  The results agree at the sub-percent level.}
\begin{center}
\begin{tabular}{|l|c|c|}
\hline
&BAO `wiggles only'& DETFast\\
\hline 
\hline
$\Omega_m h^2$&0.0261&0.0258\\
$\Omega_{\rm DE}$&0.0971&0.0967\\
$\Omega_k$&0.0308&0.0307\\
$w_0$&0.612&0.610\\
$w_p$&0.120& 0.119\\
$w_a$&2.51&2.49\\
\hline
Dark energy FoM&3.32&3.37\\
\hline
\end{tabular}
\end{center}
\label{forecasts:tab:detf}
\end{table}%

\subsection{Dependence of Figure of Merit on wavenumber range}\label{forecasts:kmax}

We find the FoM can depend strongly on the choice of non-linear cut-off or $k$-range.  We investigate this dependence by considering three different maximum wavenumbers (keeping $k_{\rm min}$ fixed at $10^{-3} h{\rm Mpc}^{-1}$), namely $k_{\max}(z)\cdot 0.5$, $k_{\rm max}(z)\cdot 1.2$ and the previous value of $k_{\rm max}(z)$.  These three wavenumber ranges evolve with redshift and have been plotted in Figure \ref{implementation:fig:kmax}.

\begin{figure}
\includegraphics[width=10cm]{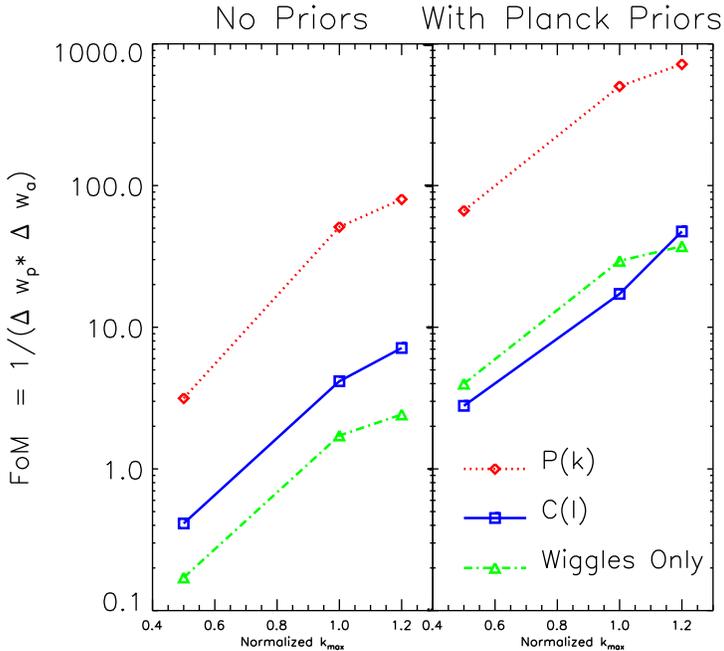}
\caption{Figure of Merit (FoM) as a function of normalized $k_{\rm max}$ for the full power spectrum $P(k)$ [dotted, diamonds], the spherical harmonic $C(\ell)$ [solid, squares] and the `wiggles only' [dot-dashed, triangles] methods - for the space based survey BAO-IVS-o.  The normalized values of $k_{\rm max}$ are taken from Figure \ref{implementation:fig:kmax}.  The left-hand panel corresponds to constraints from the BAO methods without any priors, the right-hand panel includes CMB constraints from Planck.}
\label{forecasts:fig:fom_kmax}
\end{figure}
\begin{figure}
\includegraphics[width=10cm]{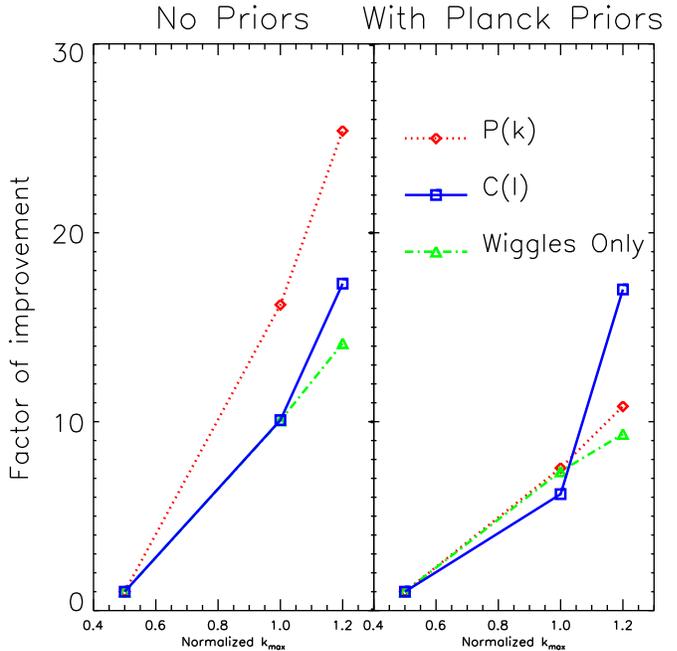}
\caption{Factor of improvement of the Figure of Merit (FoM) as a function of normalized $k_{\rm max}$ for the full power spectrum $P(k)$ [dotted, diamonds], the spherical harmonic $C(\ell)$ [solid, squares] and the `wiggles only' [dot-dashed, triangles] methods - for the space based survey BAO-IVS-o.  The improvement is quantified relatively to the value of the FoM at $k^{\rm Norm}_{\rm max} = 0.5\cdot k_{\rm max}^{\rm fid}$. The normalized values of $k_{\rm max}$ are taken from Figure \ref{implementation:fig:kmax}.  The left-hand panel corresponds to constraints from the BAO methods without any priors, the right-hand panel includes CMB constraints from Planck.  In the left panel (without Planck priors), it is clear that the FoM for the $P(k)$ methods evolves more rapidly with $k_{\rm max}$ than the $C(\ell)$ method, which in turn evolves more rapidly than the `wiggles only' method.  When Planck priors are included, the FoM from the $C(\ell)$ method evolves more rapidly than the other two methods. This may be do to different parameter degeneracies.  In this case, the FoM from the $P(k)$ method evolves more rapidly than the that for the `wiggles only' method at high $k_{\rm max}$.}
\label{forecasts:fig:norm_fom_kmax}
\end{figure}

The evolution of the FoM is shown for the three methods considered (full $P(k)$, spherical harmonics $C(\ell)$ and the `wiggles only') for the space-based survey (BAO-IVS-o) in Figures \ref{forecasts:fig:fom_kmax} and \ref{forecasts:fig:norm_fom_kmax}.  The left hand panels in both show the results without any priors and the right hand panels include Planck priors.  The FoM for the $P(k)$ method evolves rapidly with $k_{\rm max}$, jumping from $3.2$ at $k_{\rm max}(z)\cdot 0.5$ to $51$ at $k_{\rm max}^{\rm fid}(z)$ and $80$ at $k_{\rm max}(z)\cdot 1.2$.  The evolution of the $C(\ell)$ method is smoother: from 0.4 at $k_{\rm max}(z)\cdot 0.5$, to 4.2 at $k_{\rm max}^{\rm fid}(z)$, and 7.1 at $k_{\rm max}(z)\cdot 1.2$.  Finally for the `wiggles only' method, the evolution is slowest: from 0.2 at $k_{\rm max}(z)\cdot 0.5$, to 1.7 at $k_{\rm max}^{\rm fid}(z)$ and 2.4 at $k_{\rm max}(z)\cdot 1.2$.  In Figure \ref{forecasts:fig:norm_fom_kmax}, we normalize the FoM to its value at $k_{max}(z)\cdot 0.5$ in order to quantify the factor of improvement of the FoM as a function of $k_{\rm max}$.  This is plotted in Figure \ref{forecasts:fig:norm_fom_kmax}, with the left panel corresponding to results without any priors, and the right panel to results including the Planck CMB priors.

Using the normalized values, we clearly see the relative trends for each BAO method as a function of $k_{\rm max}(z)$.  Whether we consider the results with no prior or those including the Planck CMB priors, both the $P(k)$ and $C(\ell)$ methods evolve more rapidly than the `wiggles only' method, and the difference is especially noticeable as we include larger values of $k_{\rm max}$.

The full power spectrum $P(k)$ method contains information about the broad band power spectrum, and the amount of information used from the broad band power will depend on the maximum wavenumber included in the Fisher matrix calculation.

\section{Discussion}\label{discuss}

The BAO features of the galaxy power spectrum $P(k)$ have recently been observed in the latest galaxy surveys \citep{Eisenstein:2005,Hutsi:2006bao,Percival:2007b,Percival:2007}.  They are also predicted to become a fundamental tool for precision cosmology in the future \citep{WGFC, DETF}. However, there exists a range of BAO implementations in the literature.  In this paper we discuss the relative information provided by the different methods, and quantify the potential of each method to constrain dark energy parameters.

The three BAO methods we focus on are: 
\begin{itemize}
\item Full Fourier space galaxy correlation function $P(k)$ (see \cite{Seo:2003Eis}),
\item Fourier space BAO `wiggles only' (see \cite{Seo:2007Eis, Blake:2006bao}),
\item Spherical harmonic space galaxy correlation function $C(\ell)$ (see \cite{Dolney:2006}).
\end{itemize}

We do not consider the real space [$\xi(r)$] or configuration space [$w(\theta)$], correlation functions where, although the BAO signal is larger, the non-linear and linear scales are more difficult to separate.

Each method probes a different combination of features in the galaxy distribution and these are summarised in Table \ref{recipes:tab:summary}.  The full Fourier space power spectrum method probes all of these features and is thus expected to provide tighter constraints than both the `wiggles only' and $C(\ell)$ method.  The `wiggles only' method probes both the radial and tangential BAO scales - where the $C(\ell)$ method only includes information from the tangential scale (this is because the $C(\ell)$ method uses the projected 2-dimensional power spectrum).  However the $C(\ell)$ method includes information from the broad band power of the projected galaxy power spectrum.

We use the Fisher matrix formalism to forecast constraints for two future fiducial surveys, which we choose as Stage III (ground-based) and Stage IV (space-based) surveys as described in \cite[DETF]{DETF}.  The specific DETF denominations for these surveys are BAO-IIIS-o (wide survey only) and BAO-IVS-o.  We compare constraints for a 7 parameter model which allows for spatial curvature.  The evolution with redshift of the dark energy equation of state $w(z)$ is parameterised as in equation \ref{implementation:eq:wo_wa}.  Our central fiducial model in given in Table \ref{implementation:tab:central}.  The survey and implementation details are given in Table \ref{implementation:tab:stagesIII_IV}.  The choice of non-linear cut-off is justified in section \ref{implementation:details}, and its redshift dependence shown in Figure \ref{implementation:fig:kmax}.

In Table \ref{forecasts:tab:detf} we compare results from DETFast\footnote{http://www.physics.ucdavis.edu/DETFast/ by Jason Dick and Lloyd Knox.} with our calculations and find agreement at the sub percent level, using the same fiducial parameter set and values as the DETF.

The constraints for each method are presented in Table \ref{forecasts:tab:constraints}, without any priors and we also combine these results with priors from the future CMB mission Planck (the Planck priors for the same cosmological model).

The results in Table \ref{forecasts:tab:constraints} show that there exists a hierarchy between the three different methods considered here.  The $P(k)$ method which uses information from each feature in the linear galaxy power spectrum gives the tightest constraints for both the ground-based and space-based survey.  

The `wiggles only' and $C(\ell)$ methods use different information in the galaxy power spectrum.  The `wiggles only' method uses both the radial and tangential modes, but does not use information from the broad band shape of the power spectrum, the redshift distortions, nor the amplitude of the baryon wiggles.  On the other hand, the $C(\ell)$ method does not use the radial scale of the BAO scale (this is because the $C(\ell)$ method consists of the projected power spectrum).  One could use the 3-dimensional decomposition of the galaxy field in spherical harmonic space in order to recover the radial BAO information as well.

Because of this, the hierarchy between the `wiggles only' and the $C(\ell)$ method cannot be predicted.  Here we find that for the two surveys considered, the constraints for the `wiggles only' method are weaker than those from the $C(\ell)$ method, themselves are weaker than constraints from the $P(k)$ method.  For the space based survey, the Figure of Merit (FoM, defined in equation \ref{implementation:eq:fom}) for the $P(k)$ method is 35 times greater than that for the `wiggles only' method and 12 times greater than for the $C(\ell)$ method.  

We also tested how the FoM for the $P(k)$ method changed when a baryon-free spectrum was used.  The FoM dropped by a factor of 4, but was still an order of magnitude larger than the FoM for the `wiggles only' method, suggesting that a large part of the constraining power of the $P(k)$ method comes from information contained in the broad band spectrum.

Adding the Planck CMB priors to those from each BAO method reduces differences in the hierarchy.  In particular `wiggles only' constraints with Planck priors are the same order of magnitude as those for the $C(\ell)$ method with Planck priors.  In all cases constraints from the $P(k)$ method are tighter (with or without CMB priors).

We find the constraints can depend strongly on the chosen value of the non-linear cut-off, and so quantify this dependence for three choices of $k_{\rm max}(z)$.  The evolution of the FoM (with and without Planck priors) with $k_{\rm max}(z)$ is plotted in Figure \ref{forecasts:fig:fom_kmax}.  We find that there also exists a hierarchy in the evolution of the FoM with the non-linear cut-off $k_{\rm max}(z)$.  In the case where we consider the FoM from the three different methods without external priors, we find that the FoM for the full power spectrum $P(k)$ method depends strongly on $k_{\rm max}$, jumping from 3.2 to 80 for the extreme values of $k_{\rm max}$.  The dependence of the FoM for the spherical harmonic method $C(\ell)$ is weaker, from 0.4 to 7.1 over the same wavenumber range; finally, for the `wiggles only' method, the dependence is the weakest: from 0.2 to 2.4 for the same wavenumber range.  

This suggests indeed that the `wiggles only' method presents a more robust analysis, in that it depends weakly on the choice of non-linear cut-off and the wavenumber range. Both the $P(k)$ and $C(\ell)$ methods depend strongly on the wavenumber range, and will also depend on the linear bias prescription.   In any case, when discussing constraints from a BAO method, it is necessary to explicitly state the method used, as we find the FoM can vary by up to a factor of 35 between different methods.

The main conclusions of this paper are therefore:
\begin{itemize}
\item There is not a unique BAO method, but there exists a range of implementations.
\item For the three BAO implementations considered in this paper, we find that for a given fiducial survey and cosmology the figure of merit can vary by up to a factor of 35 between methods.
\item There exists a hierarchy in the constraining power of the methods. We find the more information used by a method, the higher its statistical performance, but the higher its sensitivity to systematics and implementations details.
\end{itemize}

\section*{Acknowledgments}
ARa thanks Ofer Lahav and Filipe Abdalla for useful discussions on redshift distortions as well as Sarah Bridle.  JW thanks Filipe Abdalla, Eiichiro Komatsu, Antony Lewis and Jiayu Tang for useful discussion regarding the Planck CMB prior calculation.  MK thanks Chris Blake, David Parkinson and the Fisher4Cast team for interesting discussions and help with the BAO code.  The authors also thank the EUCLID cosmology working group members for useful discussions.

\bibliographystyle{mn2e}
\bibliography{../references}

\appendix

\section{Details on Fisher Forecasts}\label{app:forecast}

It is possible to forecast the precision with which a future experiment
will be able to constrain cosmological parameters, by using the Fisher
Information Matrix (for a detailed derivation of the following see
\citealp*{Tegmark:1997th} or \citealp{Dodelson:2003}) .  This method requires only three fundamental ingredients:

\begin{itemize}
\item A set of cosmological parameters $\vec{\theta}$  for which one
  wants to forecast errors and a fiducial central model.  The parameters and fiducial values we have chosen are described in the previous section.
\item A set of $n$ measurements of the data $\vec{x}=(x_1, x_2,~...~, x_n)$ (for e.g., the spherical harmonic galaxy-galaxy power
  spectrum $C(\ell)$ over a range $\ell=1~...~n$), and a model for how the data depend on
  cosmological parameters, i.e.:
  $\vec{x}=\vec{x}\left(\vec{\theta}~\right)$
\item An estimate of the uncertainty on the data $\Delta(\vec{x})$, which may depend on the given experiment
  (instrument noise, shot noise, etc...) as well on the data estimator
  (e.g., cosmic variance).

\end{itemize}

The Fisher Information Matrix (FIM) is defined by:
\be F_{ij}=\left< \frac{\partial ^2 \mathcal{L}}{\partial \theta_i
  \partial \theta_j}\right>~,\ee
where $\mathcal{L}=-\rm{ln}$$L$ and $L=L(\vec{x},
\vec{\theta})$ is the likelihood function or the
probability distribution of the data $\vec{x}$, which
depends on some model parameter set $\vec{\theta}$. 

The forecasted uncertainty on the parameter $\theta_i$ can be estimated directly
from the FIM and obeys:\be \Delta \theta_i\ge \sqrt{(F^{-1})_{ii}}~.\ee

By assuming the errors on the estimator $\vec{x}$ are Gaussian and independent, then the Fisher matrix can be estimated by:
\be F_{ij} \simeq \sum_n \frac{1}{(\Delta x_n)^2}\frac{\partial
    x_n}{\partial \theta_i}\frac{\partial x_n}{\partial
  \theta_j} ~.\ee

The above equation holds for a single redshift; the expressions for the tomographic calculation are given below for each BAO method.

\subsection{Full Fourier space power spectrum $P(k)$}\label{app:forecast:pk}
For the case of the full Fourier space power spectrum $P(k)$, the Fisher matrix can be approximated as \citep*{Tegmark:1997th,Seo:2003Eis}:
\bea F_{ij}&=\int_{\vec{k}_{\rm{min}}}^{\vec{k}_{\rm{max}}}\frac{\partial \ln P(\vec{k})}{\partial \theta_i}\frac{\partial \ln P(\vec{k})}{\partial \theta_j}V_{\rm{eff}}(\vec{k})\frac{d\vec{k}}{2(2\pi)^3}&\\
 &=\int_{-1}^{+1}\int_{k_{\rm{min}}}^{k_{\rm{max}}}\frac{\partial \ln P(k,\mu)}{\partial \theta_i}\frac{\partial \ln P(k,\mu)}{\partial \theta_j}V_{\rm{eff}}(k,\mu)\frac{2\pi k^2dkd\mu}{2(2\pi)^3}&,\label{app:forecasts:pk:eq:fisher}\eea
where the effective volume is given by:
\be V_{\rm{eff}}(k,\mu)=\left[\frac{n(\vec{r})P(k,\mu)}{n(\vec{r})P(k,\mu)+1}\right],\ee
and the matter power spectrum is calculated using the CMBFast code \citep{CMBFast}.

Uncertainty is the redshift determination will leave tangential modes unaffected and smear radial modes.  This loss of information is modelled by \citep{Seo:2003Eis}:
\be P(\vec{k}) = P_{\rm obs}(\vec{k}) e^{-k_{\parallel}^2\sigma_r^2},\ee
i.e. by convolving the radial position by a Gaussian uncertainty.

\subsection{Fourier space BAO `Wiggles Only'}\label{app:forecast:bao}

In the case of the `wiggles only' method, only the size of the radial and tangential BAO scale is used.  This method omits all information from the shape and amplitude of the power spectrum, the redshift distortion and even the amplitude of the BAO wiggles.

We follow the method of \cite{Seo:2007Eis}, which we briefly summarise here. The starting point is the expression to compute the Fisher matrix from the full Fourier space power spectrum (equation \ref{app:forecasts:pk:eq:fisher}).  The non-linear power spectrum used is extracted from the linear power spectrum with a damping term representing the loss of information coming from the non-linear evolutionary behaviour. This damping term is approximated by an exponential computed using the Lagrangian displacement fields \cite{Eisenstein:2006}:

\be P_{\rm nl}(k,\mu)=P_{\rm lin}(k,\mu)\exp(-k^2 \Sigma^2_{\rm nl}), \ee

This non-linear damping can be decomposed into the radial and
perpendicular directions by:

\be P_{\rm nl}(k,\mu)=P_{\rm lin}(k,\mu)\exp\left(-\frac{k^2_{\perp}\Sigma^2_{\perp}}{2}-\frac{k^2_{\parallel}\Sigma^2_{\parallel}}{2}\right).\ee

A delta function baryonic peak in the correlation function at the sound horizon scale, $s$, translates into a `wiggles only' power spectrum of the form

 \be P_b \propto \frac{\sin (ks)}{ks}.\ee

This functional form is also obtained if the power spectrum is divided by a smooth version of it or if we only use the baryonic part of the power spectrum transfer function (e.g., \cite{Eisenstein:1999hu}). However, the baryonic peak is widened due to Silk damping and non-linear effects. This Gaussian broadening translates into an exponential decaying factor in Fourier space:

\bea P_b &\propto& \frac{\sin (ks)}{ks}\; \exp \left(\frac{-k^2\Sigma^2}{2}\right), \\&=& \frac{\sin (ks)}{ks}\; \exp \left[ -(k\Sigma_{\rm Silk})^{1.4}\right] \exp \left(\frac{-k^2\Sigma^2_{\rm nl}}{2}\right).\eea

We can further refine the calculation to 2 dimensions, separating the radial and tangential location of the baryon acoustic peak in the correlation function.  Using the same notation as \cite{Blake:2006bao} and \cite{Parkinson:2007}, the observables are $y(z)=r(z)/s$ and $y'(z)=r'(z)/s$  respectively, where $r(z)$ is the comoving distance to redshift z.  Measuring the fractional errors on these two observables is equivalent to measuring the fractional errors on $H\cdot s$ and $D_A/s$.

We can now use the derivatives of the `wiggles only' power spectrum by these quantities (see \cite{Seo:2007Eis} for all the details) to compute the Fisher matrix (equation 26 of \cite{Seo:2007Eis}). This equation includes the degradation due to redshift distortions in the factor $R(\mu)= (1+\beta\mu^2)^2$ - though does not include any information from these distortions (see Table \ref{recipes:tab:summary}).

The effect of photometric redshift errors can be included in this formalism as an additional exponential term in the redshift distortion factor, $R(\mu)= (1+\beta\mu^2)^2 \; \exp (-k^2\mu^2\Sigma^2_z)$, where $\Sigma_z$ is the uncertainty in the determination of photometric redshifts. We evaluate these integrals to obtain the Fisher matrix for the angular diameter distance, $D_A/s$, and the rate of expansion, $H\cdot s$.

To compute the errors on the parameters, we follow \cite{Parkinson:2007}, with some changes: Since we know the correlation between
the errors on the vector $y_i=\{r(z_i)/s, r'(z_i)/s\}$,
we form a small 2x2 covariance matrix for each redshift bin, which
we invert to obtain the corresponding Fisher matrix, $F^{(i)}$. While \cite{Parkinson:2007} compute the transformation to the cosmological parameters
analytically, we perform it numerically. 

The above is enough to provide constraints on cosmological parameters.  We also use another equivalent implementation which adds another step and is described hereafter: We form a Gaussian likelihood $L \propto \exp(-\chi^2/2)$ with
\be \chi^2 = \sum_i \sum_{a,b=1}^2 \Delta_a F_{ab}^{(i)} \Delta_b \ee
where $i$ runs over the redshift bins and $a,b$ over $r/s$ and $r'/s$,
and $\Delta=X_{\rm data}(z_i)-X_{\rm theory}(z_i)$ is the difference between the data and
the theory. We then compute numerically the matrix of second derivatives
of $\chi^2$ at the
peak of the likelihood,  via finite differencing in
all parameters (where it is necessary to divide by 2 when using
$\chi^2$ rather than $-\ln L$).

\subsection{Spherical Harmonic space $C(\ell)$}\label{app:forecast:cl}
In spherical harmonic space the estimator is the galaxy-galaxy angular 2-point function $C(\ell)$.  The Fisher matrix is then calculated using: \be F_{ij} \simeq \sum_\ell \frac{1}{\Delta C_{gg}^2(\ell)}\frac{\partial C_{gg}(\ell)}{\partial \theta_i}\frac{\partial C_{gg}(\ell)}{\partial \theta_j},\ee where \be \Delta C_{gg}(\ell) = \sqrt{\frac{2}{(2\ell+1)f_{\rm{sky}}}}\left[C_{gg}(\ell)+\mathcal{N}_{\rm g}\right], \ee and $\mathcal{N}_{\rm g}$ is the galaxy shot noise.

The galaxy 2-point correlation function in spherical harmonic space is given by: 
\be C_{gg}(\ell)=\left<|a_{\ell m}|^2\right>=4\pi b^2(z_i)\int \rd k \frac{\Delta^2(k)}{k}|W^r_\ell(k)|^2, \label{app:forecast:cl:eq:cl}\ee
where $a_{\ell m}$ represent the spherical harmonic coefficients and the galaxy bias is taken to be constant across the depth of each redshift bin.  The power spectrum can be expressed as: 
\be \Delta^2(k) = \frac{4\pi}{(2\pi)^3}k^3P(k), \ee and is calculated using the publicly available code CAMB \citep{Lewis:1999bs}. 

The real space window function $W^r_\ell(k)$ is given by: 
\be W^r_\ell(k)=\int \rd r \Theta(r)j_\ell(kr)D(r).\ee
The normalized galaxy distribution is denoted by $\Theta(r)$, the term $j_\ell(r)$ refers to the spherical Bessel function of order $\ell$ and $D(r)$ is the growth function.

In redshift space (see section \ref{clustering:zspace}) the 2-point correlation function has an extra term: 
\be C(\ell)=4\pi b^2\int \rd k \frac{\Delta^2(k)}{k}|W^r(k)+\beta W^z(k)|^2, \label{app:forecast:cl:eq:clredshift}\ee
where $\beta = \frac{\rd ln D}{\rd ln a}$ is the distortion term which modulated the amplitude of the distortion and 
\be W^z(k) = \frac{1}{k} \int \rd r \frac{d \Theta(r)}{\rd r}j'_\ell(kr).\ee The reader is referred to \cite{Fisher:1994} for a derivation.

\cite{Padmanabhan:2006} showed that this could be re-written:
\be  W^z(k)=\int
\rd r \Theta(r)\left(A_{\ell}j_\ell(kr)-B_\ell j_{\ell-2}(kr)-D_\ell j_{\ell+2}(kr)\right)\label{app:forecast:cl:eq:pad}~,\ee
where:
\be A_\ell = \frac{(2\ell^2+2\ell-1)}{(2\ell+3)(2\ell-1)}~,\ee
\be B_\ell = \frac{\ell(\ell-1)}{(2\ell+1)(2\ell-1)}~,\ee
\be D_\ell = \frac{(\ell+1)(\ell+2)}{(2\ell+1)(2\ell+3)}~.\ee

The total galaxy window including redshift distortions can then be rewritten: 
\bea W^{r+z}_{\ell}(k)=W^r(k) +\beta W^z(k)~,&&\\
=W^r_\ell(k)+\beta \left(A_\ell W^r_\ell(k)+B_\ell
W^r_{\ell-2}(k)+D_\ell W^r_{\ell+2}(k)\right) &&\label{app:forecast:eq:wz}~.\eea

So that the 2-point function with redshift distortions:
\be C(\ell)=4\pi b^2\int
dk\frac{\Delta^2(k)}{k}|W^{r+z}_\ell(k)|^2\label{app:forecast:eq:clz}~.\ee

\section{Calculation of Planck priors}\label{app:planck}

The constraints from the baryon wiggles (alone or including the full galaxy correlation) can be combined with constraints from other probes.  In this paper
we exploit measurements of the CMB anisotropies as a complementary probe; combined with constraints from the galaxy survey, this will tighten constraints on some of the cosmological
parameters. 

The primary constraint on cosmology from the CMB comes
from the measurement of the angular size of the sound horizon at last
scattering.  We will use the forthcoming Planck mission as benchmark
for a CMB prior. In order to forecast the ability of Planck to
constrain cosmological parameters we will need to estimate the errors
on the measurements of the temperature and polarization power
spectra. We will conservatively not include any B-modes in our
forecasts and assume in our fiducial model no tensor mode contribution
to the power spectra.

The Fisher matrix for CMB power spectrum is given by \citep{Zalda:1997,Zalda:1997b}:
\begin{equation}
  F_{ij}^{CMB}=\sum_{l}\sum_{X,Y}\frac{\partial
    C_{X,l}}{\partial\theta_{i}}\mathrm{COV^{-1}_{XY}}\frac{\partial
    C_{Y,l}}{\partial\theta_{j}},
  \label{eqn:cmbfisher}
\end{equation} where $\theta_i$ are the parameters to constrain, $C_{X,l}$ is the harmonic power spectrum for the temperature-temperature ($X\equiv TT$), temperature-E-polarization ($X\equiv TE$) and the E-polarization-E-polarization ($X\equiv EE$) power spectrum. The covariance
$\rm{COV}^{-1}_{XY}$ of the errors for the various power spectra is given by the fourth moment of the distribution, which under Gaussian assumptions is entirely given in terms of the $C_{X,l}$ with 

\begin{eqnarray}
{\rm COV}_{T,T} & = & f_\ell\left(C_{T,l}+W_T^{-1}B_l^{-2}\right)^2 \\
{\rm COV}_{E,E} & = & f_\ell\left(C_{E,l}+W_P^{-1}B_l^{-2}\right)^2  \\
{\rm COV}_{TE,TE} & = & f_\ell\Big[C_{TE,l}^2+\\
&&\left(C_{T,l}+W_T^{-1}B_l^{-2}\right)\left(C_{E,l}+W_P^{-1}B_l^{-2}\right)\Big] \nonumber \\
{\rm COV}_{T,E} & = & f_\ell C_{TE,l}^2  \\
{\rm COV}_{T,TE} & = & f_\ell C_{TE,l}\left(C_{T,l}+W_T^{-1}B_l^{-2}\right) \\
{\rm COV}_{E,TE} & = & f_\ell C_{TE,l}\left(C_{E,l}+W_P^{-1}B_l^{-2}\right)\; ,
\end{eqnarray}
where $f_\ell = \frac{\ell}{(2\ell+1)f_{\rm sky}}$ and $W_{T,P}=(\sigma_{T,P}\theta_{\rm fwhm})^{-2}$  is the weight per solid angle for temperature and polarization, with a $1\sigma$ sensitivity per pixel of $\sigma_{T,P}$ with a beam of $\theta_{\rm fwhm}$ extend. The beam window function is given in terms of the full width half maximum (fwhm) beam width by $B_\ell = \exp\left(-\ell(\ell+1)\theta_{\rm fwhm}^2/16\ln2\right)$ and $f_{\rm sky}$ is the sky fraction. Note that equation ~\ref{eqn:cmbfisher} usually includes a summation over the Planck frequency channels. However we conservatively assume that we will only use the 143 GHz channel as science channel, with the other frequencies used for foreground removal, which is not treated in this paper. This channel has a beam of $\theta_{\rm fwhm}=7.1'$ and sensitivities of $\sigma_T = 2.2 \mu K/K$ and $\sigma_P = 4.2\mu K/K$ \citep{Planck}.  To account for Galactic obstruction, we take $f_{\rm sky} = 0.80$. Note we use as a minimum $\ell$-mode, $\ell_{\rm min}=30$ in order to avoid problems with polarization foregrounds and subtleties for the modelling of the integrated Sachs-Wolfe effect, which depends on the specific dark energy model \citep{Weller:2003l,Caldwell:2005}. 

We have now all the ingredients to calculate the Fisher matrix forecast for Planck CMB observations. However, we still have to specify the most suitable parameter set for doing this. It is well known that one of the primary parameters which can be constrained by the CMB anisotropies is the angular size of the sound horizon \citep{Kosowsky:2002}. Also we should keep in mind that the Fisher matrix approach is a Gaussian approximation of the true underlying likelihood. In this context we would like to choose parameters in which the likelihood is as similar as possible to a Gaussian. Since the CMB anisotropies, with the exception of the integrated Sachs-Wolfe effect, are not able to constrain the equation of state of dark energy it is likely that if we would choose $(w_0,w_a)$ as parameters for the CMB Fisher matrix this degeneracy is artificially broken by the Fisher matrix approach. This was also recognized by the DETF and we hence follow their approach to calculate the CMB Fisher matrix \citep{DETF}. 

We choose as fiducial parameter set $\vec{\theta}= (\omega_m, \theta_S,\ln A_S, \omega_b, n_S, \tau)$, where $\theta_S$ is the angular size of the sound horizon at last scattering \citep{Kosowsky:2002}, $\ln A_S$ is the logarithm of the primordial amplitude of scalar perturbations and $\tau$ is the optical depth due to reionization. Note that there might be an even more suitable parameter set, which is currently explored \citep{Mukherjee:2008}. The optical depth in terms of the BAO analysis presented here is a nuisance parameter and we marginalise over it analytically. We then calculate the Planck CMB Fisher matrix with the help of the publicly available CAMB\footnote{http://camb.info} code \citep{Lewis:1999bs}. As a final step we transform the Planck Fisher matrix in the DETF parameter set to the one used in the analysis presented in this paper. The parameters here are given by ${\tilde \theta} = (\Omega_m, \Omega_{de},h, \sigma_8,  \Omega_b, w_0, w_a, n_S)$ by using the transformation with  the Jacobian
\be
J_{\tilde \alpha \alpha} = \frac{\partial
  \theta_\alpha}{\partial\tilde\theta_{\tilde\alpha}}
\ee
and the Fisher matrix in our set of basis parameters
\be
\tilde{\mathbf{F}} = {\mathbf J}{\mathbf F}{\mathbf J}^T\;.
\ee

\bsp

\label{lastpage}

\end{document}